# Approximate Solution to the Fractional Second Type Lane-Emden Equation

Emad A-B. Abdel-Salam[1,2] and M. I. Nouh[3,4]

[1]Department of Mathematics, Faculty of Science, Northern Border University, Arar 1321, Saudi Arabia.
[2]Department of Mathematics, Faculty of Science, Assiut University, New Valley Branch, El-Kharja 72511, Egypt.
Email:emad_abdelsalam@yahoo.com.
[3]Department of Physics, Faculty of Science, Northern Border University, Arar 1321, Saudi Arabia.
[4]Department of Astronomy, National Research Institute of Astronomy and Geophysics, 11421 Helwan, Cairo, Egypt

**Abstract.**

The spherical isothermal Lane-Emden equation is a second order non-linear differential equation that model many configurations in astrophysics. In the present paper and based on the fractal index technique and the series expansion, the fractional lane-Emden equation involving modified Riemann-Liouville derivative is solved. The results indicate that, the series converges for the radius range $0 \leq x < 2200$ with fractional parameter $\alpha$ spreads over a wide range of the fractional parameter $\alpha$. Comparison with the numerical solution revealed a good agreement with a maximum relative error 0.05.

1. Introduction

The isothermal Lane–Emden equation is often considered as the asymptotic limit of the Lane–Emden equation where the polytropic index is taken to be very large. Self-gravitational isothermal gas sphere have been useful in many areas of astrophysics, such as stellar structure, star clusters, galaxies and galactic clusters, Binnyand Tremaine (1987) and Horedt (2004).

Many numerical and analytical methods have been proposed for the solution of the isothermal gas spheres, of them are; Natarajan & Lynden-Bell (1997), Roxburgh& Stockman (1999), Hunter (2001), Nouh (2004), Mirza (2009) and Soliman and Al-Zeghhayer (2015).

In the last decades, there has been a surge of interest in studying fractional differential equations (FDEs) which describes many branches of science such as mathematics, chemistry, optics, plasmas, fluid dynamics and engineering. Applications of fractional calculus and fractional differential equations examples include: dielectric relaxation phenomena in polymeric materials (Miller and Ross, 1993), transport of passive tracers carried by fluid flow in a porous medium in groundwater hydrology (Zaslavsky, 2005), transport dynamics in systems governed by anomalous diffusion (Hilfer (2000), West (2000)), long-time memory in financial time series (Diethelm, 2010) and so on. So it is very important to find efficient methods for solving fractional differential equations. Finding analytical and approximate solutions of FDEs is one of the most useful approaches to understand the physical mechanism of natural phenomenon and dynamically processes modeled by FDEs (Abdel-Salam (2013), (2014), (2015a), (2015b)).

In the present paper, we introduce new analytical solution to the isothermal gas sphere. We construct a recurrence relation for the coefficients in the power series expansion of the solution of the fractional isothermal Lane–Emden equation. With the best to our knowledge, this is the first work deals with the series solution of the fractional isothermal Lane-Emden equation.

The structure of the paper is as follows. In section 2, some basic concepts of the fractional calculus are introduced. The series solution to the fractional isothermal gas sphere is described in section 3. Section 4 is devoted to the numerical results. Section 4 deals with the conclusion reached.

2. **Basics of the Fractional Calculus**

Fractional calculus is the generalizations of ordinary calculus. There are many kinds of fractional calculus. Such as Riemann–Liouville, Caputo, Kolwankar–Gangal, Oldham and Spanier, Miller and Ross, Cresson's, Grunwald-Letnikov, and modified Riemann–Liouville (Mandelbrot (1982), Kolwankar (1996) and Kolwankar (1997)).

We start by remembering the Jumarie modification of Riemann–Liouville derivative, Jumarie (2009), (2010), (2012a), (2012b), (2013). Assume that $f : R \to R, x \to f(x)$ denote a continuous function, and let $h$ denotes a constant discretization span, the limit form of the modified Riemann–Liouville derivative

$$f^{(\alpha)}(x) = \lim_{h \downarrow 0} \frac{\Delta^\alpha[f(x)-f(0)]}{h^\alpha}, \qquad 0 < \alpha < 1,$$

where $\Delta^\alpha f(x) = \sum_{k=0}^{\infty}(-1)^k \frac{\Gamma(\alpha+1)}{\Gamma(k+1)\Gamma(\alpha-k+1)} f[x+(\alpha-k)h]$. Which is similar to the standard of derivatives (calculus for beginners), and the $\alpha$-order derivative of a constant is zero. The integral form of the modified Riemann–Liouville derivative is written as

$$D_x^\alpha f(x) = \begin{cases} \dfrac{1}{\Gamma(-\alpha)} \displaystyle\int_0^x (x-\xi)^{-\alpha-1}[f(\xi)-f(0)]d\xi, & \alpha < 0 \\[2mm] \dfrac{1}{\Gamma(1-\alpha)} \dfrac{d}{dx} \displaystyle\int_0^x (x-\xi)^{-\alpha}[f(\xi)-f(0)]d\xi, & 0 < \alpha < 1 \\[2mm] \dfrac{1}{\Gamma(n-\alpha)} \dfrac{d^n}{dx^n} \displaystyle\int_0^x (x-\xi)^{n-\alpha-1}[f(\xi)-f(0)]d\xi, & n \le \alpha < n+1, n \ge 1. \end{cases} \quad (1)$$

Another some useful Jumarie modified formulae could be written as

$$D_x^\alpha x^\gamma = \frac{\Gamma(\gamma+1)}{\Gamma(\gamma+1-\alpha)} x^{\gamma-\alpha}, \qquad \gamma > 0, \tag{2}$$

$$D_x^\alpha (c f(x)) = c D_x^\alpha f(x), \tag{3}$$

$$D_x^\alpha [f(x)g(x)] = g(x) D_x^\alpha f(x) + f(x) D_x^\alpha g(x), \tag{4}$$

$$D_x^\alpha f[g(x)] = f_g'[g(x)] D_x^\alpha g(x), \tag{5}$$

$$D_x^\alpha f[g(x)] = D_g^\alpha f[g(x)](g_x')^\alpha, \tag{6}$$

where $c$ is constant. Equations (4) to (6) are direct results from

$$D_x^\alpha f(x) \cong \Gamma(\alpha+1) D_x f(x). \tag{7}$$

He et al. (2012) modified the chain rule equation (5) to the formula

$$D_x^\alpha f[g(x)] = \sigma_x f_g'[g(x)] D_x^\alpha g(x), \tag{8}$$

where $\sigma_x$ is called the fractal index which is usually determined in terms of gamma functions Ibrahim (2012), He et al. (2012) and Abdel-Salam et al (2015c). Therefore, Equations (4) and (6) will be modified to the following forms

$$D_x^\alpha [f(x)g(x)] = \sigma_x \{g(x) D_x^\alpha f(x) + f(x) D_x^\alpha g(x)\}, \tag{9}$$

$$D_x^\alpha f[g(x)] = \sigma_x D_g^\alpha f[g(x)](g_x')^\alpha. \tag{10}$$

Throughout this manuscript, we use Equation (8) to solve the fractional isothermal Lane Emden equation.

### 3. Computational Developments
#### 3.1. Isothermal Lane Emden Equation

The Lane–Emden equation for an isothermal gas sphere, Binney and Tremaine (1987), can be written as

$$\frac{d^2 u}{dx^2} + \frac{2}{x}\frac{du}{dx} = e^{-u} \tag{11}$$

with the initial condition

$$u(0) = 0, \qquad \frac{du}{dx}\bigg|_{x=0} = 0 \tag{12}$$

The series solution of Equation (11) have the form (Nouh, 2004)

$$u(x) = \frac{x^2}{6} - \frac{x^4}{120} + \frac{x^6}{1590} - \frac{61 x^8}{1632960} + - \ldots . \tag{13}$$

The fractional isothermal Lane-Emden equation which is the generalization form of the isothermal Lane-Emden equation (11) could be written as:

$$x^{2\alpha} D_x^\alpha D_x^\alpha u + 2 x^\alpha D_x^\alpha u + x^{2\alpha} e^{-u} = 0, \tag{14}$$

with the initial conditions

$$u(0) = 0, \qquad D_x^\alpha u(0) = 0, \tag{15}$$

where $u = u(x)$, is an unknown function, $D_x^\alpha$ is the modified Riemann–Liouville derivative and

$$e^{-u} = \sum_{s=0}^{\infty} \frac{(-1)^k u^k}{k!}.$$

#### 3.2. Series Solution

We assume the transform $X = x^\alpha$ and the solution could be expressed in a series form as:

$$u(X) = \sum_{m=0}^{\infty} A_m X^m = A_0 + A_1 X + A_2 X^2 + A_3 X^3 + A_4 X^4 + A_5 X^5 + \ldots$$
$$= A_0 + A_1 x^\alpha + A_2 x^{2\alpha} + A_3 x^{3\alpha} + A_4 x^{4\alpha} + A_5 x^{5\alpha} + \ldots \qquad (16)$$

The first initial condition of Equation (16) gives $u(0) = A_0$, or $A_0 = 0$.

Applying Equation (2) and Equation (4) to Equation (16) we found

$$D_x^\alpha u = D_x^\alpha A_0 + D_x^\alpha (A_1 x^\alpha) + D_x^\alpha (A_2 x^{2\alpha}) + D_x^\alpha (A_3 x^{3\alpha}) + D_x^\alpha (A_4 x^{4\alpha}) + D_x^\alpha (A_5 x^{5\alpha}) + \ldots$$
$$= 0 + \frac{A_1 \Gamma(\alpha+1) x^{\alpha-\alpha}}{\Gamma(\alpha+1-\alpha)} + \frac{A_2 \Gamma(2\alpha+1) x^{2\alpha-\alpha}}{\Gamma(2\alpha+1-\alpha)} + \frac{A_3 \Gamma(3\alpha+1) x^{3\alpha-\alpha}}{\Gamma(3\alpha+1-\alpha)} + \ldots \qquad (17)$$
$$= A_1 \Gamma(\alpha+1) + \frac{A_2 \Gamma(2\alpha+1) x^\alpha}{\Gamma(\alpha+1)} + \frac{A_3 \Gamma(3\alpha+1) x^{2\alpha}}{\Gamma(2\alpha+1)} + \ldots$$

Applying the second initial condition of Equation (12) gives

$$D_x^\alpha u(0) = A_1 \Gamma(\alpha+1), \quad \text{or} \quad A_1 = 0. \qquad (18)$$

Now suppose that

$$G(X) = \sum_{m=0}^{\infty} Q_m X^m = Q_0 + Q_1 X + Q_2 X^2 + Q_3 X^3 + Q_4 X^4 + Q_5 X^5 + \ldots \qquad (19)$$

By putting

$$e^{-u} = G(X), \qquad (20)$$

we have

$$E_\alpha(-u(0)) = G(0) = 1, \quad \text{or} \quad Q_0 = 1.$$

First, we adapt our self to find the fractional derivative of $e^{-u} = \sum_{s=0}^{\infty} \frac{(-1)^k u^k}{k!}$, taking the fractional derivative for the both sides and note that the fractional derivative of $u^2$, it will be considered as $u$ times $u$ similarly $u^3$ will be considered as $u$ times $u^2$ and so on, therefore

$$D_x^\alpha e^{-u} = D_x^\alpha \left( \sum_{k=0}^{\infty} \frac{(-1)^k u^k}{k!} \right) = \sum_{k=1}^{\infty} \frac{(-1)^k D_x^\alpha u^k}{k!} = \sum_{k=0}^{\infty} \frac{(-1)^k k u^{k-1} D_x^\alpha u}{k!}$$
$$= \sum_{k=1}^{\infty} \frac{(-1)^k u^{k-1} D_x^\alpha u}{(k-1)!} = -D_x^\alpha u \sum_{k=1}^{\infty} \frac{(-1)^{k-1} u^{k-1}}{(k-1)!} = -D_x^\alpha u \sum_{s=0}^{\infty} \frac{(-1)^s u^s}{s!} = -e^{-u} D_x^\alpha u,$$

this can be written in the form

$$GD_x^\alpha u = -D_x^\alpha G. \tag{21}$$

Differentiating both sides of Equation (21) $k$ times $\alpha$ − derivaives we have

$$\underbrace{D_x^\alpha ... D_x^\alpha}_{k\,times}[GD_x^\alpha u] = -\underbrace{D_x^\alpha ... D_x^\alpha}_{k\,times}(D_x^\alpha G), \quad \text{or} \quad \sum_{j=0}^{k}\binom{k}{j}u^{\alpha(j+1)}G^{\alpha(k-j)} = -G^{\alpha(k+1)}$$

at $x = 0$

$$\sum_{j=0}^{k}\binom{k}{j}u^{\alpha(j+1)}(0)G^{\alpha(k-j)}(0) = -G^{\alpha(k+1)}(0) \tag{22}$$

since

$$u^{\alpha(j+1)}(0) = A_{j+1}\Gamma((j+1)\alpha+1),\ G^{\alpha(k-j)}(0) = Q_{k-j}\Gamma((k-j)\alpha+1),\ G^{\alpha(k+1)}(0) = Q_{k+1}\Gamma((k+1)\alpha+1),$$

we have

$$\sum_{j=0}^{k}\frac{k!A_{j+1}\Gamma((j+1)\alpha+1)Q_{k-j}\Gamma((k-j)\alpha+1)}{j!(k-j)!} = -Q_{k+1}\Gamma((k+1)\alpha+1)$$

that is

$$Q_{k+1} = -\sum_{j=0}^{k}\frac{k!A_{j+1}\Gamma((j+1)\alpha+1)Q_{k-j}\Gamma((k-j)\alpha+1)}{j!(k-j)!\Gamma((k+1)\alpha+1)},$$

let $l = k+1$ then

$$Q_l = -\frac{(l-1)!}{\Gamma(l\alpha+1)}\sum_{j=0}^{l-1}\frac{A_{j+1}\Gamma((j+1)\alpha+1)Q_{l-1-j}\Gamma((l-1-j)\alpha+1)}{j!(l-1-j)!},$$

If $i = j+1$

$$Q_l = -\frac{(l-1)!}{\Gamma(l\alpha+1)}\sum_{i=1}^{l}\frac{A_i\Gamma(i\alpha+1)Q_{l-i}\Gamma((l-i)\alpha+1)}{(i-1)!(l-i)!}, \tag{23}$$

That's gives

$$A_0 = 0, \quad A_1 = 1, \quad Q_0 = 1, \quad Q_1 = 0. \tag{24}$$

Now $D_x^\alpha u$ could be written as

$$D_x^\alpha u = \sum_{m=1}^{\infty} A_m \sigma_x\, mX^{m-1} \frac{\Gamma(\alpha+1)}{\Gamma(\alpha+1-\alpha)} x^{\alpha-\alpha}$$

$$= \sum_{m=1}^{\infty} A_m mX^{m-1} \frac{\Gamma(\alpha+1)}{\Gamma(\alpha+1-\alpha)} \frac{\Gamma(m\alpha+1)}{m\Gamma(\alpha+1)\Gamma(m\alpha+1-\alpha)} \quad (25)$$

$$= \sum_{m=1}^{\infty} A_m X^{m-1} \frac{\Gamma(m\alpha+1)}{\Gamma(m\alpha+1-\alpha)},$$

where the fractal index $\sigma_x$ (see example 6 in Abdel-Salam et al (2015)) is given by

$$\sigma_x = \frac{\Gamma(m\alpha+1)}{m\Gamma(\alpha+1)\Gamma(m\alpha+1-\alpha)},$$

Also, $D_x^\alpha D_x^\alpha u$ could be written as

$$D_x^\alpha D_x^\alpha u = \sum_{m=2}^{\infty} A_m \sigma_x (m-1)X^{m-2} \frac{\Gamma(m\alpha+1)}{\Gamma(m\alpha+1-\alpha)} \frac{\Gamma(\alpha+1)}{\Gamma(\alpha+1-\alpha)} x^{\alpha-\alpha},$$

$$= \sum_{m=2}^{\infty} X^{m-2} \frac{A_m(m-1)\Gamma(m\alpha+1)}{\Gamma(m\alpha+1-\alpha)} \frac{\Gamma(\alpha+1)}{\Gamma(\alpha+1-\alpha)} \frac{\Gamma((m-1)\alpha+1)}{(m-1)\Gamma(\alpha+1)\Gamma((m-1)\alpha+1-\alpha)} \quad (26)$$

$$= \sum_{m=2}^{\infty} X^{m-2} \frac{A_m \Gamma(m\alpha+1)}{\Gamma(m\alpha+1-2\alpha)}$$

with the fractal index $\sigma_x$ (see example 6 in Abdel-Salam et al (2015)) is given by

$$\sigma_x = \frac{\Gamma((m-1)\alpha+1)}{(m-1)\Gamma(\alpha+1)\Gamma((m-1)\alpha+1-\alpha)}$$

Substituting Equations (19), (25) and (26) into Equation (14) yields

$$x^{2\alpha} \sum_{m=2}^{\infty} X^{m-2} \frac{A_m \Gamma(m\alpha+1)}{\Gamma(m\alpha+1-2\alpha)} + 2x^\alpha \sum_{m=2}^{\infty} X^{m-1} \frac{A_m \Gamma(m\alpha+1)}{\Gamma(m\alpha+1-\alpha)} - x^{2\alpha}\left[1+\sum_{m=1}^{\infty} Q_m X^m\right] = 0,$$

$$\sum_{m=2}^{\infty} X^m \frac{A_m \Gamma(m\alpha+1)}{\Gamma(m\alpha+1-2\alpha)} + \sum_{m=2}^{\infty} X^m \frac{2A_m \Gamma(m\alpha+1)}{\Gamma(m\alpha+1-\alpha)} - \left[X^2 + \sum_{m=1}^{\infty} Q_m X^{m+2}\right] = 0, \quad (27)$$

$$\sum_{m=0}^{\infty} X^{m+2} \frac{A_{m+2}\Gamma((m+2)\alpha+1)}{\Gamma(m\alpha+1)} + \sum_{m=0}^{\infty} X^{m+2} \frac{2A_{m+2}\Gamma((m+2)\alpha+1)}{\Gamma(m\alpha+1+\alpha)} - \left[X^2 + \sum_{m=1}^{\infty} Q_m X^{m+2}\right] = 0.$$

Equating the coefficients of $X^2$ and $X^{m+2}$ we get

$$A_2 = \frac{\Gamma(\alpha+1)}{\Gamma(2\alpha+1)[\Gamma(\alpha+1)+2]}, \quad (28)$$

and

$$A_{m+2} = \frac{\Gamma(m\alpha+1)\,\Gamma((m+1)\alpha+1)}{\Gamma((m+1)\alpha+1)\,\Gamma((m+2)\alpha+1)+2\Gamma(m\alpha+1)\,\Gamma((m+2)\alpha+1)} Q_m. \quad (29)$$

Equations (23) and (29) are the recurrence relations of the series expansion, Equation (16).

## 4. Results

We elaborated a FORTRAN code to calculate the series coefficients for the range $0 \leq x < 2200$, this range is that of an isothermal sphere which is on the brink of gravothermal collapse, Hunter (2001). We ran the code with a step $\Delta\alpha = 0.05$ to visit large range of the fractional parameters $\alpha$ and varying the number of series terms till we obtain the minimum value of the relative error.

Figure (1) plots the Emden function $u$ versus $x$. Figure (2) shows the variation of the relative error with $x$, the maximum relative error is about 0.05 which represent good accuracy. The variation of $\alpha$, is plotted in Figure (3). As we notice, generally as $x$ increases the $\alpha$, parameter decreases except for some intermediate values. This result shows the highly dependence of the convergence on $\alpha$. The number of series terms required to obtain a solution with suitable relative error is lower than that of the accelerated series solution of Nouh (2004) by more than 50%.

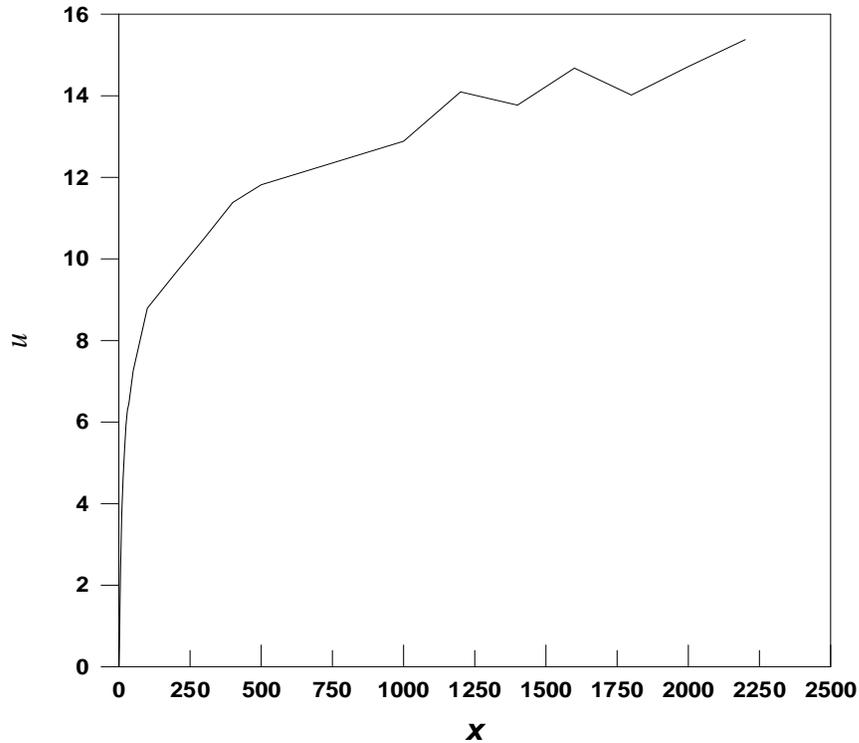

Figure (1): The Emden function $u$ versus the radius of the isothermal gas sphere $x$.

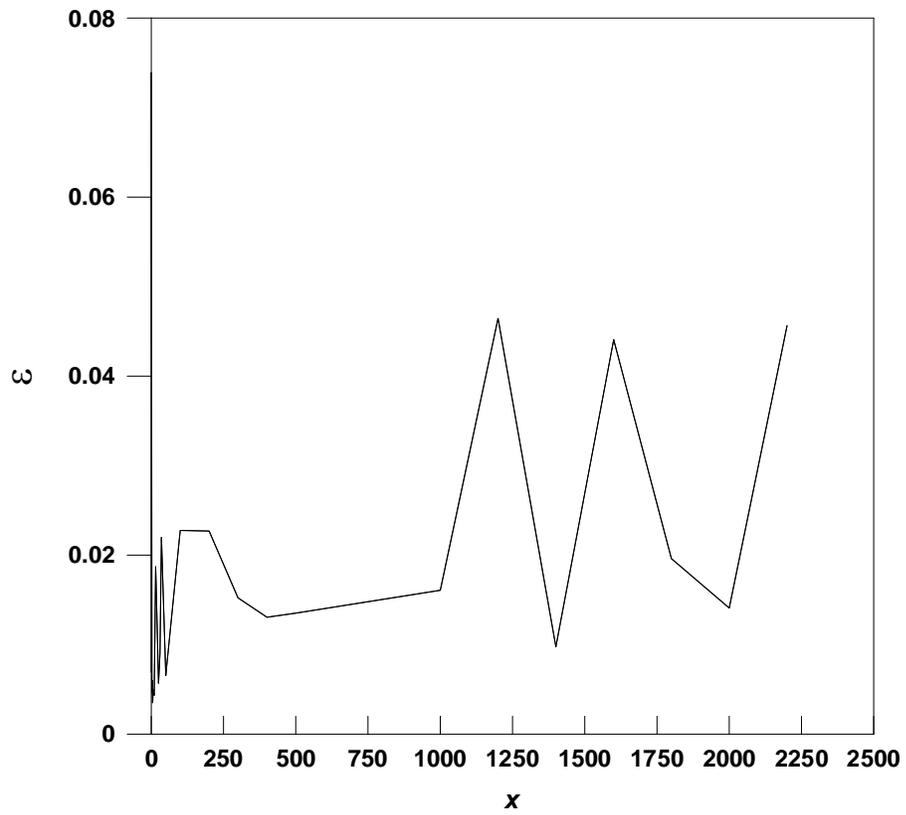

Figure (2): Absolute relative errors in the Emden function u.

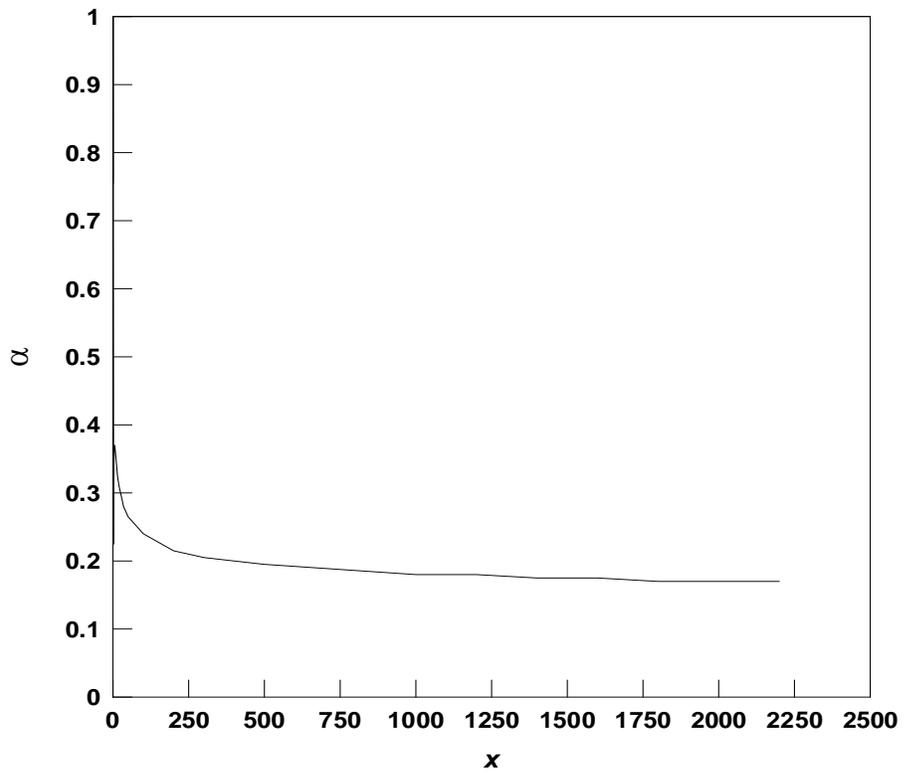

Figure (3): Variation of the fractional parameter $\alpha$ over the radius of the isothermal gas sphere.

## 5. Conclusion

We derived a series solution for the fractional isothermal Lane Emden equation. Two recurrence relations are derived and solved simultaneously. What is called the fractal index is deduced for each term in the series expansion.

We ran our code for a small step on fractional parameter $\alpha$ to declare the effects of this factor on the accuracy of the calculations. The series reaches the surface of the sphere faster when applying $\alpha$, which may be considered as an accelerator of the series. We found that the range of $\alpha$ spreads on all the range $0 < \alpha \leq 1$. The results show that, the maximum relative error is 0.05 which reflect a good agreement with the numerical values. Application of the procedure on the first type Lane Emden equation (polytropic gas sphere) will be done in the near future.

**Acknowledgements**

We wish to acknowledge the financial support of this work through Northern Border University, deanship of scientific research and higher education grant number 5-27-1436-5.

**Appendix A:**

In this appendix we shall determine some series coefficients in the series expansion of Equation (16).

By putting $l = 2$ in Equation (23) and using Equation (28) for $A_2$ we get

$$\begin{aligned} Q_2 &= -\frac{(2-1)!}{\Gamma(2\alpha+1)} \sum_{i=1}^{2} \frac{A_i \Gamma(i\alpha+1) Q_{2-i} \Gamma((2-i)\alpha+1)}{(i-1)!(2-i)!} \\ &= -\frac{1}{\Gamma(2\alpha+1)} \left[ \frac{A_1 \Gamma(\alpha+1) Q_1 \Gamma(\alpha+1)}{0!(1)!} + \frac{A_2 \Gamma(2\alpha+1) Q_0 \Gamma(1)}{1!(0)!} \right], \\ &= -\frac{\Gamma(\alpha+1)}{\Gamma(2\alpha+1)[\Gamma(\alpha+1)+2]} \end{aligned} \qquad (A1)$$

Put $m = 1$ in Equation (29) we have

$$A_3 = \frac{\Gamma(\alpha+1)\,\Gamma(2\alpha+1)}{\Gamma(2\alpha+1)\,\Gamma(3\alpha+1) + 2\Gamma(\alpha+1)\,\Gamma(3\alpha+1)} Q_1 \tag{A2}$$

since $Q_1 = 0$, it follows that $A_3 = 0$.

Again use Equation (23) with $l = 3$ we get

$$\begin{aligned}Q_3 &= -\frac{2!}{\Gamma(3\alpha+1)} \sum_{j=0}^{2} \frac{A_{j+1}\Gamma((j+1)\alpha+1) Q_{2-j} \Gamma((2-j)\alpha+1)}{j!(2-j)!} \\ &= -\frac{2!}{\Gamma(3\alpha+1)} \left[ \frac{A_1 \Gamma(\alpha+1) Q_2 \Gamma((2\alpha+1)}{2!} + \frac{A_2 \Gamma(2\alpha+1) Q_1 \Gamma(\alpha+1)}{1!1!} + \frac{A_3 \Gamma(3\alpha+1) Q_0}{2!} \right] = 0\end{aligned} \tag{A3}$$

To calculate $A_4$, we put $m = 2$ in Equation (29) and using Equation (A1) for $Q_2$ gives

$$A_4 = -\frac{\Gamma(\alpha+1)\Gamma(2\alpha+1)\Gamma(3\alpha+1)}{\Gamma(4\alpha+1)\Gamma(2\alpha+1)\left[\Gamma(3\alpha+1)+2\Gamma(2\alpha+1)\right]\left[\Gamma(\alpha+1)+2\right]}, \tag{A4}$$

Proceed as above we get

$$Q_4 = \frac{\Gamma(\alpha+1)}{\Gamma(4\alpha+1)[\Gamma(\alpha+1)+2]} \left[ \frac{3\Gamma(\alpha+1)}{[\Gamma(\alpha+1)+2]} + \frac{\Gamma(3\alpha+1)}{[\Gamma(3\alpha+1)+2\Gamma(2\alpha+1)]} \right], \tag{A5}$$

$A_5 = 0,$

$Q_5 = 0$

$$A_6 = \frac{\Gamma(\alpha+1)\Gamma(5\alpha+1)}{\Gamma(6\alpha+1)[\Gamma(\alpha+1)+2][\Gamma(5\alpha+1)+2\Gamma(4\alpha+1)]} \left[ \frac{3\Gamma(\alpha+1)}{\Gamma(\alpha+1)+2} + \frac{\Gamma(3\alpha+1)}{\Gamma(3\alpha+1)+2\Gamma(2\alpha+1)} \right] \tag{A6}$$

Now put $\alpha = 1$ in Equations (A1) to (A6) we get the coefficient of the series expansion (Equation (16)) as follows

$$A_0 = 0,\ A_1 = 0, A_2 = \frac{1}{6},\ A_3 = 0,\ A_4 = -\frac{1}{120},\ A_5 = 0,\ A_6 = \frac{1}{1890}$$

Which is the same as the series solution of Equation (12).